\documentclass[preprint,preprintnumbers,aps,amsmath,amssymb,superscriptaddress]{revtex4} 

\usepackage{graphicx}

\newcommand{\qed}{\hfill $\Box$}

\newcommand{\bN}{\mathbb N}

\newcommand{\nin}{\notin}

\newcommand{\cR}{\mathcal R}

\newcommand{\fut}{\mathrm{Fut}}
\newcommand{\past}{\mathrm{Past}}
\newcommand{\vol}{\mathrm{vol}}

\newcommand{\cD}{\mathcal D}

\newcommand{\union}{\bigcup}
\newcommand{\cQ}{\mathcal Q}

\newcommand{\stem}{\mathrm{stem}}
\def\be{\begin{displaymath}}
\def\ee{\end{displaymath}}
\def\bne{\begin{equation}}
\def\ene{\end{equation}}
\def\leqsim{\stackrel{\scriptstyle <}{\scriptstyle \sim}}
\newcommand{\maxfutend}{maximal future end}
\newcommand{\ta}{thickened antichain} 
\newcommand{\tas}{thickened antichains}
\newcommand{\ia}{inextendible antichain}

\begin{document}

\title{Spatial Hypersurfaces in Causal Set Cosmology}

\author{Seth A. Major} 
\email{smajor@hamilton.edu}
\affiliation{Department of Physics, Hamilton College, Clinton NY 13323
USA}  
\author{David Rideout} 
\email{d.rideout@imperial.ac.uk}
\affiliation{Blackett Laboratory, Imperial College, London SW7 2AZ, UK} 
\author{Sumati Surya} \email{ssurya@rri.res.in}
\affiliation{Raman Research Institute, CV Raman Avenue, Sadashivanagar,
Bangalore - 560 080, INDIA}

\date{January 2006}

\begin{abstract} 
Within the causal set approach to quantum gravity, a discrete analog
of a spacelike region is a set of unrelated elements, or an
``antichain''. In the continuum approximation of the theory, a
moment-of-time hypersurface is well represented by an ``inextendible''
antichain.  We construct a richer structure corresponding to a
``thickening'' of this antichain containing non-trivial geometric and
topological information.  We find that covariant observables can be
associated with such \tas{} and transitions between them, in classical
stochastic growth models of causal sets.  This construction highlights
the difference between the covariant measure on causal set cosmology
and the standard sum-over-histories approach: the measure is assigned
to completed histories rather than to histories on a restricted spacetime
region. The resulting re-phrasing of the sum-over-histories may be fruitful
in other approaches to quantum gravity.
\end{abstract}

\maketitle 

\section{Introduction}

A major hindrance to the development of quantum gravity is the family
of issues known as the problem of time \cite{IK}.  Simply put, the
problem is one of labeling states to record the evolution of physical
quantities. In the canonical formulation, the problem of time arises
because the Hamiltonian, which encodes the dynamics, is a constraint;
since the observables commute with the constraints, they become
``frozen''.  Thus, it appears that there are no dynamical observables
in the theory
\footnote{However, observables may be expressed as correlations
between dynamical degrees of freedom.  See \cite{CR} and references
therein.}. In the path integral formulation, the problem arises both
in finding a covariant way of expressing the initial and final states
of geometry and in finding a measure such that the propagator sums
over physically distinct configurations.

One approach to this problem is to abandon all reference to spatial
geometries and instead consider only spacetime quantities.  However,
since much of classical and quantum physics is conceptualized in terms
of moment-of-time and initial value formulations, it is important to
know whether a covariant interpretation of spatial hypersurfaces is at
all possible.  The Hartle-Hawking no-boundary proposal, for example,
provides one such interpretation within the framework of Euclidean
sum-over-histories quantum gravity \cite{HH}.

We examine this question within the causal set approach to quantum
gravity. Our analysis suggests a general prescription for other
sum-over-histories approaches to quantum gravity: instead of summing
over histories between geometries at fixed times the measure is on
``completed'' histories containing transitions between the spatial
hypersurfaces. This may be particularly relevant to the Causal
Dynamical Triangulation (CDT) approach to quantising gravity in which
one considers ``sufficiently'' completed histories \cite{loll}.

In causal set quantum gravity, the underlying structure is a causal
set $C$, which is a locally finite partial order \cite{causets,RSlec}:
i.e., $C$ is a set with an order relation $\prec$ which is transitive
($x \prec y$ and $ y \prec z \Rightarrow x \prec z$), irreflexive ($ x
\nprec x $) and locally finite ($|\past(x) \cap \fut(y)| < \infty$)
for any $x,y,z \in C$ (where $\past(y)\!=\! \{x| x\prec y \}$,
$\fut(y)\!=\!\{x| y \prec x \}$ and $|A|$ denotes set-cardinality.)
The spacetime continuum arises as an approximation, rather than a
limit of the theory, and quantities such as geometry and topology
correspond to macroscopic degrees of freedom, analogous to
thermodynamic quantities such as pressure and temperature for a gas of
particles. While we expect a causal set to include both microscopic
and macroscopic physical information, the latter should be
well-described by physics in the continuum approximation.

A continuum manifold $(M,g)$ is said to approximate a causal set $C$
if and only if there exists a faithful embedding $\phi:C \rightarrow
M$.  The faithful embedding requires that the causal relations induced
by the embedding agree with the order relation in $C$, and that the
embedded points are sprinkled randomly via a Poisson process, at a
spacetime density $\rho$, with high probability. A causal set embedded
in a spacetime $(M,g)$ thus resembles a random spacetime lattice,
where a link between two points implies a causal relation. In this
lattice, a set of unrelated elements corresponds to a set of mutually
spacelike events.

In the continuum, we define a {\sl moment-of-time surface (MoTS)} to
be a spacelike, acausal, edgeless hypersurface such that any other
spacetime event is either to its causal future or past. This
definition captures our requirement that the spatial hypersurface
represent a global moment-of-time. An example of a MoTS is a Cauchy
hypersurface in a globally hyperbolic spacetime.
 In the discrete context of causal sets, an {\sl antichain} is a set of
unrelated elements and corresponds, in an embedding, to a set of
spacelike events.  Thus, a natural
 analog of a MoTS is an {\sl inextendible antichain}, i.e.\ an
antichain such that any element not in it is related to it
 \footnote{ An analogy between a spatial hypersurface and an 
 inextendible antichain (often called a maximal antichain) is 
suggested in \cite{rds}. We clarify that the appropriate continuum 
analog to an inextendible antichain is a MoTS.}. 

How good is this analogy? Given a faithful embedding $\phi:C
\rightarrow M$, if $A$ is an inextendible antichain in $C$, there
exist an uncountable infinity of MoTSs which contain
$\phi(A)$.  To what degree do they differ?
Let $\Sigma_{1,2}$ be two such MoTSs.
Consider any pair of points $p_{1,2}$ on $\Sigma_{1,2}$ for which the
spacetime interval $I^+(p_1) \cap I^-(p_2)$ is non-empty
(where $I^\pm(x)$ represents the chronological future/past of
$x$).
Since $\phi(A)$ is inextendible any such interval does not intersect
$\phi(C)$.  The probability for this to have happened via a Poisson
sprinkling is given by $\exp[-\rho
\, {\vol}(I^+(p_1) \cap I^-(p_2))]$, which is appreciable only for
${\vol}(I^+(p_1) \cap I^-(p_2)) \leqsim 1/\rho $. Thus, with ``high
probability''
$\phi(A)$ represents a MoTS up to $\rho^{-1}$ differences, i.e. a MoTS
is a good approximation to $A$. The analogy with a MoTS is further
strengthened by results in \cite{nerves} in which we show that the
homology of a MoTS can be recovered from an \ia{} which it contains.
However arising  from the Lorentz invariance of the sprinkled
causet, a crucial difference
persists.  Unlike a MoTS, $\phi(A)$ need not ``capture'' all irreducible
causal relations or links, i.e., if $x,y \in C$ such that $x\prec y$,  $x \in
\past(A)$, $y \in \fut(A)$, and if there exists no $z \in A$ for
which $x\prec z \prec y$.  The existence of such
``missing links'' might lead to observational consequences.
 
Since an antichain is simply a collection of unrelated elements, it is
characterized only by its cardinality, which is not enough information
to record dimension and topology, much less geometry.  We therefore
need a richer structure to extract the relevant information contained
in a MoTS from the causal set.  Indeed, as shown in
\cite{stanley,finitary,graham}, the larger causal set contains
geometric and topological information.  One might expect that
not all the information in the causal set is required for this
purpose, but rather only an appropriately defined subcauset which is a
neighbourhood of the inextendible antichain.  Such a neighbourhood has
a continuum interpretation as a ``thickened'' MoTS with the scale of
thickening corresponding roughly to the resolution of time in physical
measurements.

\begin{figure}
       \begin{center}
   \begin{tabular*}{\textwidth}
{c@{\extracolsep{\fill}}c@{\extracolsep{\fill}}}
  \includegraphics[scale=.5]{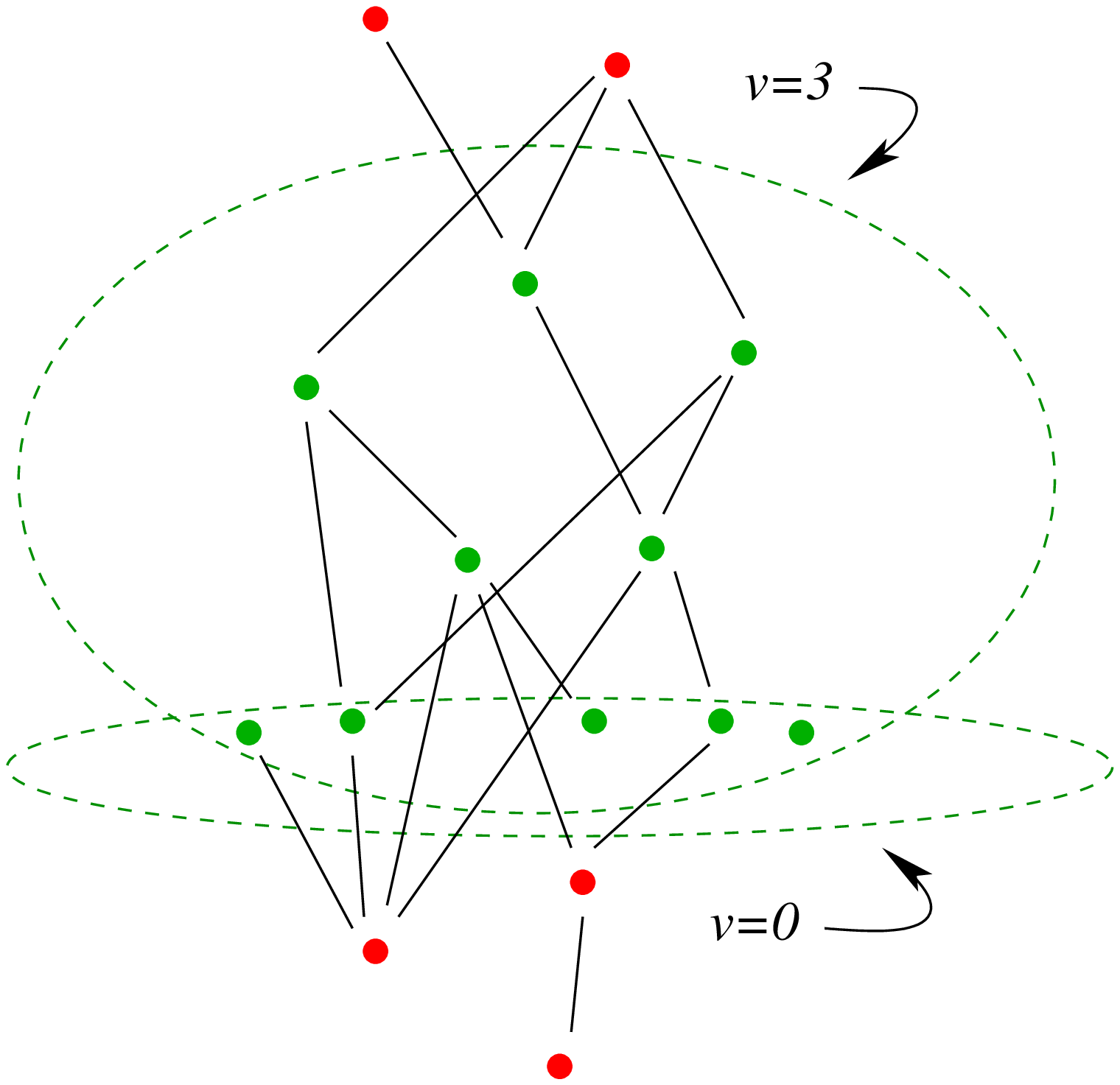}
    &
   \includegraphics [scale=1.1]{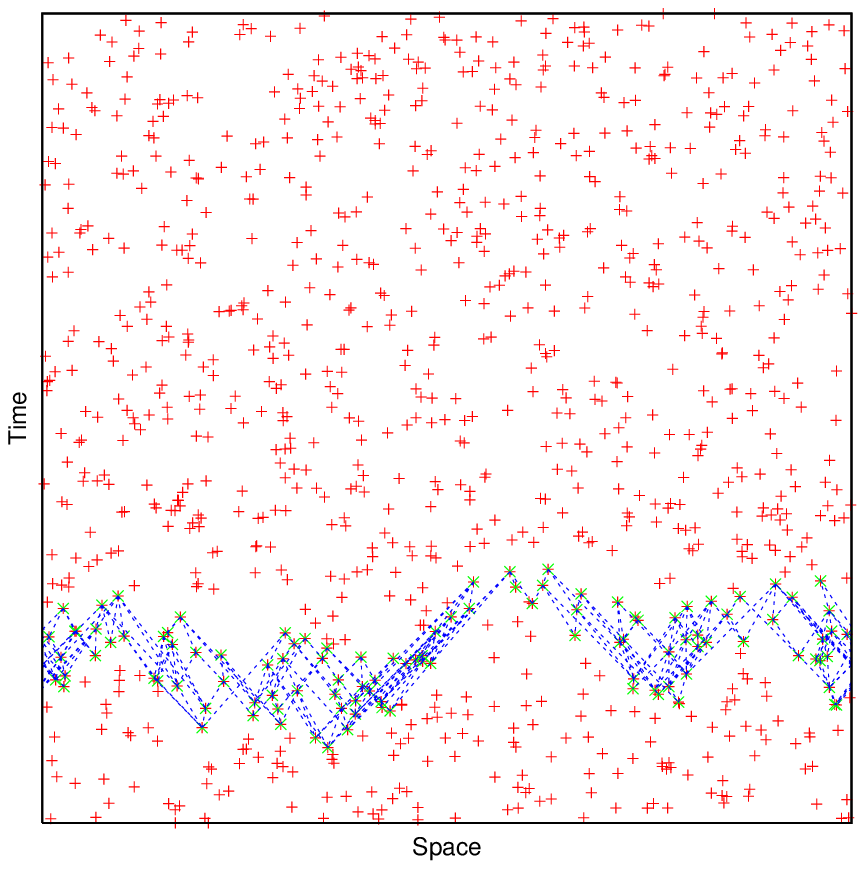} \\
   (a.) &  (b.)
   \end{tabular*}
   \end{center}
\caption{\label{figs} (a) An inextendible antichain in an abstract 
causal set is delineated by the $v=0$ dashed line. The larger $v=3$ 
thickened antichain is also delineated. Elements are shown as dots and links by solid lines.  (b) A thickened antichain
with $v=16$ is shown in a sprinkled causet embedded in $S^1 \times I$.
Elements in the sprinkling are denoted with a (red) ``+" and 
elements in the \ta{} are denoted with a (green) ``$\times$". The 
causal relations in the \ta{} are shown as  (blue) dotted lines.}
\end{figure}

One particularly useful definition of this
neighbourhood is a future or past volume thickening of the
inextendible antichain,  which can be used to recover the 
homological information of a spatial hypersurface \cite{nerves}. An
inextendible antichain $A$ is thickened by a volume $v$ by adjoining
to $A$ the set of elements to its future(past) whose past(future) up to
and including $A$ has cardinality less than or equal to $v$.  We
define a (future volume) {\sl \ta{}} to be
\begin{displaymath}
A_v = \left\{x|x\in \fut(A) \cup A \;\mathrm{and}\; |\past(x)
 \setminus \past(A)| \leq  v \right\}, 
\end{displaymath}
so that $A_0=A$ (see Fig. \ref{figs}.)  A past volume \ta{} is
similarly defined.  For both future and past volume thickenings, $A_v$
is {\sl convex} in the larger causal set $C$, i.e. if $x \prec z \prec
y $ for any $x,y \in A_v$, then $z \in A_v$.

The question that is of interest to us here is whether the discrete
analogs of a MoTS and transitions between them can be given a
covariant meaning in causal set quantum gravity.  In the standard
sum-over-histories formulation, the Hartle-Hawking wavefunction
for a spatial geometry $(\Sigma,q)$ is
\begin{equation*} 
\psi(q;\Sigma) = \sum_M \int \cD g \, e^{-S_E(g)},  
\end{equation*} 
where the sum is over manifolds $M$ with boundary $\Sigma$ and the
path integral is over (Euclidean) 
histories $g$ with $g|_\Sigma=q$. This amplitude $\psi(q;\Sigma)$ cannot be 
interpreted as the amplitude for $(\Sigma,
q)$ to be a MoTS, since this requires that all subsequent events be to
its causal future; only {\it completed} spacetime histories contain
such subsequent information.  To implement this within Lorentzian
quantum gravity one requires the subclass $\Gamma
\subset \{(M,g)\}$ of the set of  ``completed'' histories containing $(\Sigma, q)$ as a
MoTS.  The amplitude is then $\psi(q;\Sigma)= \sum_M \int_{g} \cD g e^{iS[g]}$, where
$(M,g) \in \Gamma$.

Similarly, the transition amplitude between two
spatial slices $(\Sigma_1, q_1)$ and $(\Sigma_2, q_2)$, cannot be
interpreted as a transition amplitude between MoTSs. One must
restrict to the subclass of completed histories $\Gamma_{12}$ such
that for any $(M,g) \in \Gamma_{12}$, $\Sigma_1, \Sigma_2$ are
non-intersecting MoTSs with $\Sigma_1$ to the past of $\Sigma_2$ and
with the interpolating region between them either to the past or the
future, but not both, of any other event in the spacetime.  The transition 
amplitude between the two MoTSs is then $\sum_M
\int_g \cD g e^{iS[g]}$, with $(M,g) \in \Gamma_{12}$. This
transition amplitude doesn't depend merely on
the action of the interpolating region  but also on the measure of the set of
completed histories $\Gamma_{12}$ which contains such a region.  This
is a non-local prescription and differs significantly from that in the
standard sum-over-histories approach.

Although this prescription is natural in the causal set context, it also provides 
a new framework for covariant sum-over-histories which may be useful
in other approaches to quantum gravity.  In the remainder of this article 
we briefly review the role of observables in causal set ``classical 
sequential growth" dynamics in the next section. In section \ref{proof}
we show that the existence of thickened antichains, and transitions
between them, is measurable.  They are observables.  In the final
section \ref{discussion} we related these measures to more familiar
conditional probabilities and reflect on how these ideas might be applied
in the context of causal dynamical triangulations.

\section{Observables in classical sequential growth} 

In the absence of a clear formulation of a quantum causal set
dynamics, we turn to the classical stochastic growth (CSG) models of
causets \cite{rideoutsorkin,ep} to flesh out the details of
our proposal. Covariant observables in the class of (extended)
generalised percolation models have been completely characterised
\cite{observables,oep} and the hope is that they will provide an
insight into the nature of observables in the quantum theory.  We use
the formalism developed in \cite{observables} to show  that 
 the causal set analogue of MoTSs and transitions between them can be
assigned a covariant meaning in these theories.

In \cite{observables} the set of observables for a class of CSG
models, called generalised percolation dynamics, was completely
characterised in terms of covariant past sets or ``stem sets''. These
results were recently generalised to the class of extended percolation
dynamics \cite{ep} in \cite{oep}. More specifically, one can express
the causal set dynamics in terms of a measure space, which is a triad
$(\Omega, \cR, \mu)$, with $\Omega$ the set of {\sl completed}
unlabeled causets being the sample space, $\cR$ a $\sigma$-algebra on
$\Omega$, and $\mu$ the probability measure obtained from the
dynamics.  A covariant observable of the theory then corresponds to a
measurable physical question, i.e., an element of $\cR$.

A subcauset $S$ of a causal set is called a {\sl stem} if it is its
own inclusive past, i.e., $S=\past(S)\cup S$.  A {\sl stem set}
associated with a (finite) causet $S$ is the set of completed causal
sets $C$ which contain $S$ as stem.  The main result of
\cite{observables,oep} was to show that, save sets of measure zero,
the set of stem sets generates the $\sigma$-algebra $\cR$.  Moreover,
since stems are past sets, the physical interpretation of these
observables is straightforward.  What we will prove is that \tas{} and
transitions between them admit a covariant, measurable interpretation
in terms of stem sets and hence are elements of $\cR$.

Because of the formulation of the dynamics as a sequential growth
process, we need to impose the slightly stronger condition of {\sl
past finiteness} on the causets: $C$ is past finite if $|\past(x)|$ is
finite $\forall x \in C$. Let $\Omega(\bN)$ be the set of finite
causets.  A finite causet $E$ is a {\sl future end} of a finite causet
$C$ if $E$ is a subcauset of $C$ such that $E$ contains its own future
in $C$.  The {\sl past end} is defined analogously and is equivalent
to a stem.  A future end $E$ of $C$ is said to be {\sl maximal} if for
every $e \nin E$, $e \prec E$.  A maximal past end is similarly
defined. A finite causet $I$ is said to {\sl interpolate} between the
subcausets $A$ and $B$ if $I$ has a maximal past end isomorphic to
$A$, and a maximal future end of isomorphic to $B$, both of which are
\ta{}s in $I$ and are disjoint in $I$.  A causet $C$ contains an {\sl
interpolating region} $I$ with maximal ends $\{A, B \}$ if $I$ is a
finite convex subcauset of $C$ interpolating between a maximal future
and past end isomorphic to $A$ and $B$, respectively, which are
both  thickened antichains in $C$. 

In the CSG models, one grows a causal set stage by stage, starting
from a single element. Following the terminology of that paper, a
finite causal set $P \in \Omega(n)$ is an {\sl ancestor} of some $D
\in \Omega(\bN)\sqcup \Omega$ if $P$ is a stem in $D$.  Conversely,
$D$ is said to be a {\sl descendant} of $P$.  One obtains $D$ from $P$
by adding elements to $P$ maximally: starting from $P$, a single
maximal element is added to $P$ to obtain the causal set $P'$, and so
on to get $D$. The new elements are maximal at the stage at which they
are added, but do not need to remain maximal at subsequent stages.

For simplicity, we will confine our analysis to \tas{} of finite
cardinality and interpolators between such \tas{} of fixed finite
cardinality.  A finite \ta{} can be given the continuum interpretation
of a ``volume thickened'' MoTS which has finite spacetime
volume. Likewise, fixing the cardinality of the interpolator between a
pair of finite \tas{} corresponds to fixing the volume of the
spacetime between two MoTSs, as in the unimodular approach to quantum
gravity \cite{unimodular}.

\section{Observables, thickened antichains, and transitions}
\label{proof}

Let $\Phi_A$ denote the set of (past finite) completed causets
containing the finite future or past volume thickening $A$ of an
inextendible antichain $a$. Given a pair of finite causets $\{A, B\}$
containing inextendible antichains $a$ and $b$ respectively, and an $N
\in \bN$, let $\Phi^N_{AB}$ denote the set of completed causets
containing any interpolator $I \in \Omega(N)$ with ends $\{A, B\}$ as
an interpolating region. Assuming throughout that $A$, $B$ are
non-empty, we show that:

\vskip 0.1cm
\noindent {\bf Proposition:} {\em (i) $\Phi_A$ and (ii) $\Phi^N_{AB}$
are elements of the $\sigma$-algebra $\cR(S)$ generated by the set of
stem sets. They are therefore measurable sets in the generalised and
extended percolation dynamics of CSG.} \\

\noindent {\bf Proof:} Given the result of \cite{observables,oep}, it
suffices to demonstrate that the sets $\Phi_A$ and $\Phi^N_{AB}$ lie
in the $\sigma$-algebra generated by stem sets $\cR(S)$. \\

\noindent (i) Let $\cQ$ be the set of finite causets with $A$ as a
\maxfutend. For any $P\in \cQ$, let $\{A_i\} $ be the (finite) set of
isomorphic copies of $A$ contained in $P$ as a maximal future end,
with each $A_i$ containing an inextendible antichain $a_i$ isomorphic
to $a$.  Let $S_P$ be the set of finite descendants of $P$ which do
not contain {\it any} of the $A_i$ as a \ta.  Define the sets
\begin{displaymath} 
\!\!\! \Xi_P \equiv\! \!\union_{D\in S_P} \!\! \!\stem(D), \,  \Phi_P \equiv 
 \stem(P)\backslash \Xi_P, \, \Phi \equiv \! \union_{P\in \cQ}
 \Phi_P. 
\end{displaymath} 
Clearly, $\Phi \in \cR(S)$, since $\cQ$ is countable, and $\Xi_p$ is a
countable union of stem sets for any $P\in\cQ$.  We now show that
$\Phi\subseteq \Phi_A$.  Assume otherwise, i.e., $\exists \, \,
C\in\Phi$ which does not contain a subcauset isomorphic to $A$ as a
\ta.  Since $C \in \Phi$, $ \exists \, \, P\in \cQ$ such that $C \in
\stem(P)$. $P$ contains a set of isomorphic copies $\{A_i \}$ of $A$
as maximal future ends, with inextendible antichains $a_i$, none of
which are \tas{} in $C$.  Since $C \in \stem(P)$, the $A_i$ are convex
in $C$ and the $a_i$ are antichains in $C$.  Thus the only option then
is that none of the $a_i$ are inextendible antichains in $C$. Thus,
$C$ contains a finite stem $D \supset P$ for which this is true, so
that $D \in S_P$. But since $C \in \stem(D)$, $C \in \Xi_P$ or $C\in
(\Phi_P)^c$. This must be true for all $P \in \cQ_C$ where $\cQ_C=\{ P
\in Q| C \in \stem(P)\}$. Thus, $C \in \bigcap_{P \in Q_C} (\Phi_P)^c
= (\bigcup_{P\in Q_C} \Phi_P)^c$. Since for $P \in \cQ \backslash
\cQ_C$, $C \nin \stem(P)$, this means that $C\nin\Phi$, which is a
contradiction. Next, we show that $\Phi_A \subseteq \Phi$.  Let $C \in
\Phi_A$ and take $P$ to be the inclusive past of a subcauset $A$ in
$C$ which occurs as a \ta.  $P$ is therefore a stem in $C$, so that $C
\in \stem(P)$. Since $A$ is a \ta{} in $C$, there exists no stem $D$
in $C$ containing $P$ for which the $a$ in $A$ is not 
 an \ia.  Neither can the convexity of $A$ be spoiled by any
descendant.  Thus $C \nin \Xi_P$, and hence $C \in \Phi$.  \qed
\vskip 0.1cm
\noindent {Note:} The measurability of the set of past finite causets
containing a ``post'', or an element related to every other element in
the causet, is a special case, when $A$ is a single element antichain.

\noindent (ii) The measurability of $\Phi_{AB}^N$ follows similarly.
There exist a finite set $\Psi$ of interpolators $I \in \Omega(N)$
with ends $\{ A, B\}$. Let $\cQ$ be the finite set of causets which
contain any $I_i \in \Psi$ as an interpolating region {\it and} as a
\maxfutend.  For any $P\in \cQ$, let $\{ I_{i} \}$ be the (finite
number of) $I$'s $\in \Psi$ contained in $P$ as an interpolating
region and as a \maxfutend, and let the ends of each $I_{i}$ be $\{
A_{ik}, B_{il} \}$, where $A_{ik}, \, B_{il}$ are isomorphic to $A$
and $B$, respectively, with $a_{ik}, b_{il}$ the respective
inextendible antichains isomorphic to $a,b$. Let $S_P$ be the set of
descendants of $P$ which do not contain any copy of any of the $I_i
\subset P$ as an interpolating region.  We may then define the set
$\Phi \in \cR(S)$ as in (i). The proof is now logically identical to
that of (i).\qed

The proof above can be simply modified to accommodate more generally
defined \tas.  For example, a \ta{} $A$ can be defined to be a convex
subcauset of $C$ which contains {\it any} antichain $a \subset A$
inextendible in $C$.  Or, it could be defined to be convex and
``separating'' in $C$, the latter term meaning that every element of
$C$ not in $A$ is related to an element of $A$.

\section{Discussion}
\label{discussion}

As is obvious from our construction, the measure of $\Phi_A$ is the
probability that the \ta{} $A$ appears {\it at least once} in a causal
set.  For example, if the measure of $\Phi_A$ is one, then this means
that all completed causets that can occur in the dynamics contain $A$
as a thickened antichain, and if it is zero, then none of the causets
can contain $A$ as a \ta. Similarly, the measure of $\Phi^N_{AB}$ is
the probability that an interpolating region of cardinality $N$ with
the ends $A$ and $B$ appears at least once in a causet. This provides
a covariant definition of transitions between discrete analogs of
MoTSs.  Relating the measure $\mu(\Phi^N_{AB})$ on $\Phi^N_{AB}$ to a
transition probability, however, is a more subtle issue since it is
the conditional probability for the transition from $A$ to $B$ {\it
given} that $A$ occurs \footnote{We thank Joe Henson for pointing this
out.}. If we merely divide $\mu(\Phi^N_{AB})$ by the measure
$\mu(\Phi_A)$ on $\Phi_A$, then this gives us the probability that a
causet will contain at least one interpolating region from $A$ to $B$,
given that $A$ occurs.  This differs from the standard interpretation,
since it leaves open the possibility that other $A$'s in a causet may
not transition to $B$ with the same probability.
  
The measure on $\Phi_A$ can be used to determine the measure on
spatial topologies in the following sense: spatial topologies
corresponding to $A$ are precluded if $\Phi_A$ has vanishingly small
probability measure. Likewise, the occurrence of transitions from $A$
to $B$ are precluded if $\Phi_{AB}$ has vanishingly small
probability. Restricting to a ``cosmic time'' $T$, which we take to be
the cardinality of the inclusive past, is also useful for cosmological
predictions, since we can ask what is the probability for $A$ to occur
at some suitable $T$. On classical scales, one doesn't expect to see
topology change \cite{tgychange}. Hence, if $T$ is chosen to be large
enough so that topology change can no longer occur, then this would
amount to a prediction for the spatial topology of the universe, in
the event that our analysis survives quantisation.

We have used the CSG model to test our hypothesis that \tas{} and
transitions between them can be given a covariant interpretation in
causal set dynamics. Moreover, we suggest that the continuum
approximation of a \ta{} is a thickened MoTS which contains the
geometric and topological information of a MoTS. This provides us with
a new covariant understanding of spatial hypersurfaces and transitions
between them. Indeed, if the observability of stem sets survives
quantisation, our analysis and interpretation would extend trivially
to the quantum picture.
 
What is new --- and may have relevance outside the context of causal
sets --- is that we have defined a covariant procedure for identifying
a sub-region of the universe, independent of a choice of coordinate
time. This notion of a covariant sum-over-histories may be useful in
other approaches to quantum gravity in computing transitions 
over a finite spacetime volume and on completed histories.
In CDT, since the histories are sufficiently completed, the
probabilities can be obtained by performing a search in
the space of causal triangulations for a particular three geometry
$A$, 
or for a triangulated region of volume $V$ containing $A$ and $B$
as initial and final boundaries.  Here $A$ and $B$ do not necessarily have to be individual geometries, but could be required to be members of sets of geometries which posses some property.
The notion of a thickened antichain also finds a resonance in CDT. In
\cite{loll} it was shown that the spectral dimension for spatial
hypersurfaces is 1.5 rather than 3 in 4-dimensional spacetimes and
that thickening by a single unit gives a value of a little over 2. Our
work on the homology of thickened antichains \cite{nerves} suggests
that the spatial hypersurface needs to be thickened to a scale much
larger than the discreteness scale before it can correctly reproduce
continuum topological information, including dimension.

Causal sets also make their appearance in the spin foam framework, 
as causal spin foams (CSF)\cite{Fotini}. Our prescription suggests a
way to obtain a spatial hypersurface in an abstract CSF, where again, 
the thickening roughly corresponds to a time-measurement.  To the
extent that the CSF approach is a sum-over-histories formulation of
quantum gravity, our construction suggests a way of  obtaining
covariant (spacetime) observables in this approach. 

\begin{flushleft}
We would like to thank Fay Dowker and Joe Henson for discussions. This work
was supported in part by a Research Corporation grant to SAM.
\end{flushleft}

\end{document}